\documentclass[12pt]{article}
\usepackage{epsfig} \usepackage{amssymb} \usepackage{amsfonts}
\def\eV{{\rm e\kern-0.12em V}} \def\GeV{{\rm G}\eV} \def\MeV{{\rm M}\eV}
\def\MSbar{\relax\ifmmode\overline{\rm MS}\else{$\overline{\rm MS}${ }}\fi}
\def\msbar{\relax\ifmmode\overline{\rm MS}\else{$\overline{\rm MS}${ }}\fi}
\def\alphan{\relax\ifmmode{\alpha_{\rm an}}\else{$\alpha_{\rm an}${ }}\fi}
\def\albars{\relax\ifmmode{\bar{\alpha}_s}\else{$\bar{\alpha}_s${ }}\fi}
\def\albarsQ{\relax\ifmmode{\bar{\alpha}_s(Q^2)}\else{$\bar{\alpha}_s(Q^2)${ }}\fi}
\def\asmz{\relax\ifmmode\bar \alpha_s(M_Z^2)\else{$\bar \alpha_s(M_Z^2)${ }}\fi}
\def\asmzs{\relax\ifmmode{\albars(M_z^2)}\else{{$\albars(m_z^2)$}{ }}\fi}
\def\tildal{\relax\ifmmode{\tilde{\alpha}}\else{$\tilde{\alpha}${ }}\fi}
\def\tildals{\relax\ifmmode{\tilde{\alpha}(s)}\else{$\tilde{\alpha}(s)${ }}\fi}
\def\asQ{\relax\ifmmode\bar{\alpha}_s(Q^2)\else{$\bar{\alpha}_s(Q^2)${ }}\fi}
\def\as{\relax\ifmmode\bar{\alpha}_s\else{$\bar{\alpha}_s${ }}\fi}
\def\agoth{\relax\ifmmode{\mathfrak A}\else{${\mathfrak A}${ }}\fi}
\def\pisq{\relax\ifmmode{\pi^2}\else{${\pi^2}${ }}\fi}
\def\agothk{\relax\ifmmode{\mathfrak A}_k\else{${\mathfrak A}_k${ }}\fi}
\def\acal{\relax\ifmmode{\cal A}\else{${\cal A}${ }}\fi}
\def\acalk{\relax\ifmmode{\cal A}_k\else{${\cal A}_k${ }}\fi}
\newcommand{\beglab}{\begin{equation}\label}
\newcommand{\beq}{\begin{equation}} \newcommand{\eeq}{\end{equation}}
\begin{document}
\begin{center}
{\large\sf The $\pi^2$ terms in the $s$--channel QCD observables } \\
\bigskip

D.V. Shirkov \\
{\it Bogoliubov Laboratory, JINR, 141980 Dubna, Russia\\
e-address: shirkovd@thsun1.jinr.ru} \end{center} \smallskip

\centerline{\bf Abstract} {\small We analyze the effect of
$\pi^2$--terms in the QCD perturbative expansions for the
$s$--channel effective coupling and observables, the effect known
from the 80s. We remind that these terms can be collected into
specific functions --- strong $s$--channel coupling \tildals and
its effective powers $\agothk(s)$ free of ghost singularities.
Further on, we study the structure of perturbation theory for
observables and its reformulation in terms of nonpower
perturbation expansion over the set $\left\{\agothk(s)\right\}.$\par

Then we discuss the influence of this effect on the numerical
values of \as as extracted from experiments. The main result is
that the common two-loop (NLO, NLLA) approximation widely used in
the five-quark ($10\:\,\GeV\lesssim \sqrt{s}\lesssim 170\:
\,\GeV$) region for a shape analysis contains a systematic
negative error of a 1--2 per cent order of magnitude for the
extracted $\albars^{(2)}\,.$  Our physical conclusion is that the
\asmz value averaged over the $f=5\,$ data $$<\asmz>_{f=5} \simeq
0.124\,$$ appreciably differs from the currently accepted ``world
average" $(=0.118\,)$.}\par \medskip

\section{\sf Preamble}

 Usually, physical quantities in the time-like channel, like the cross-section
ratio of the inclusive $e^+e^- \to$ hadron annihilation or the $\tau$--decay
process,  are presented in the form of two- or three-term perturbation
expansion
 \beq \label{rnew} \frac{R(s)}{R_0}=1+r(s)\,;\quad
r(s)=c_1\,\albars(s)+c_2\,\albars^2+ c_3\,\albars^3 +\dots \,\eeq
(our coefficients $c_k=C_k\,\pi^{-k}$ are normalized differently from the
commonly adopted, like in Refs.\cite{pdg00,beth00, bardin}) over powers
of effective QCD coupling \albars which is supposed {\it ad hoc\,} to be of
the same form as in the Euclidean domain, e.g.,
\begin{eqnarray}\label{als3}
\albars^{(3)}(s)&=&\frac{1}{\beta_0L}-\frac{b_1}{\beta_0^2}\frac{\ln L}
{L^2} +\frac{1}{\beta_0^3L^3}\left[b_1^2(\ln^2L-\ln L-1)+b_2\right]; \,
\nonumber \\
 &+&\frac{1}{\beta_0^4L^4}\left[b_1^3\left(-\ln^3L+\frac{5}{2}\ln^2L+
2\ln L-\frac{1}{2}\right)- 3b_1b_2\ln L+\frac{b_3}{2}\right]\,.\nonumber
\end{eqnarray}

   Here, $ L=\ln (s/\Lambda^2)\,$ and for the beta-function we use
normalization
$$\beta(\alpha)=-\beta_0\,\alpha^2-\beta_1\,\alpha^3-\beta_2\,\alpha^4+\dots
=-\beta_0\,\alpha^2\left(1+b_1\,\alpha +b_2\,\alpha^2+\dots\right)\,,$$
that is also free of $\pi$ powers. Numerically,
$$\beta_0(f)=\frac{33-2\,f}{12\pi} \,;~\  b_1(f)=\frac{153-19f}{2\pi(33-2f)}
\,;\ b_1(4\pm 1)=0.490^{-0.089}_{+0.076}\,. $$

 Coefficients $c_{k\geq 3}=d_k-\delta_k\,$ include ``$\pi^2$ structures"
$\delta_k\,$ proportional to lower $c_k$:
{\small
\begin{equation}\label{deltas}
\delta_3=\frac{(\pi\beta_0(f))^2\,c_1}{3}\,,~~\delta_4=(\pi\beta_0)^2\,(c_2+
\frac{5}{6}\:b_1\,c_1)\,; \:\ \pi^2\beta_0^2(4\pm 1)=4.340^{-.666}_{+.723}\,.\eeq}
 These structures $\delta_k\,$ arise\cite{rad82,kras82,bjork89,kat95} in the
course of analytic continuation from the Euclidean to Minkowskian region.
Coefficients $d_k\,$ should be treated as a genuine $k$th--order ones. Just
they have to be calculated with the help of relevant Feynman diagrams. \smallskip

 To illustrate, consider the three--flavor  case for $\tau$--decay,
$f=4\,,5\,$ cases for $e^+e^- \to$ hadron annihilation and $Z_0$ decay
(with $f=5$) --- see Table 1 in which we also give values for the
$\pi^2$--terms\,.\smallskip

\begin{center} {\sf\large Table 1}  \label{tab1}\smallskip

\begin{tabular}{|c|c||c|c|c|c|c|c|}  \hline
Process & f &$c_1$&$c_2=d_2$& $c_3$ &$d_3=c_3-\delta_3$&$\delta_3$&$\delta_4$
\\ \hline\hline

$\tau$ decay& 3&$1/\pi$&.526&$0.852$& 1.389& 0.537  & 5.01 \\ \hline

$e^+e^-$    & 4 &.318 &.155 &-\,0.351& 0.111 & 0.462 & 2.451 \\ \hline

$e^+e^-$    & 5 &.318 &.143 &-\,0.413 &-\,0.023& 0.390  & 1.752 \\ \hline

$Z_0$ decay & 5 &.318&.095 &-\,0.483 &-\,0.094& 0.390  & 1.576  \\ \hline
\end{tabular} \end{center}

  Here, all coefficients $\:c_k\,,\: d_k\,$ and $\:\delta_{k}\,,$ due to
normalization (\ref{rnew}), are of an order of unity. One can see that, in the
high energy region, contribution of $\:\delta_3\,$ prevails in $\:c_3\,.$

\section{\sf  Preliminary quantitative estimate \label{s2}} 
 In practice, the $\pi^2$--terms often dominate in higher expansion
coefficients. This effect is especially strong in the $f=5\,$ region.
Meanwhile, just in this region people often
use the so-called NLLA approximation, that is the two-term representation
 \beq\label{two-term}
O(s) = C_1 (\albars/\pi) +C_2 (\albars/\pi)^2 \eeq
for an observable $O(s)$ when next, the three-loop, coefficient $C_3$\, is
not known. This is the case, e.g., with event--shape\cite{delphi} analysis. \par
\smallskip

 On the basis of the numerical estimates of Table 1, in such a case, we
{\sf recommend to use the three-term expression}
\beq\label{3Delta}
O_3^{\Delta}(s)=d_1\left\{\albars-\frac{\pi^2\beta_0^2}{3}\albars^3\right\}
+d_2\albars^2 =c_1\,\albars+ c_2\albars^2- \underline{{\bf\delta_3\,
\albars^3}}\,\eeq
i.e., to take into account the known predominant \pisq part of the next
coefficient $c_3\,$. As it follows from the comparison of the last expression
with the previous, two--term one, the \albars numerical value extracted
from eq.(\ref{3Delta}), for the same measured value $O_{obs}$, will differ
by a positive quantity (e.g., in the $f=5\,$ region with $\,\albars \simeq
0.12 \div 0.15)$
$$
(\triangle\albars)_3=\left.\frac{\pi\delta_3\,\albars^3}{1+2\pi d_2
\albars}\right|_{20\div 100 \GeV}^{f=5}=\frac{1.225\,\albars^3}
{1+0.90\,\albars}\simeq0.002\div0.003\, $$
that turns to be numerically important.
\par \smallskip

  Moreover, in the $f=4\,$ region, where the three-loop approximation is
commonly used in the data analysis, the \pisq term $\delta_4\,$ of the next
order turns out also to be essential. Hence, we {\sf propose to use the
four-term expression}
\beq\label{4thterm}
O_4^{\Delta}(s)= d_1\,\albars+ d_2\,\albars^2 +c_3\,\albars^3-
\underline{\mathbf{\delta_4\,\albars^4}}\,; \quad c_3=d_3-\delta_3\,\eeq
 (instead of the three-term one (\ref{rnew})) that is equivalent to
{\small
\beq\label{4Delta}
O_4^{\Delta}(s)= d_1\left\{\albars-\frac{\pi^2\beta_0^2}{3}\albars^3
-b_1\frac{5}{6}\,\pi^2\beta_0^2\,\albars^4\right\} +d_2\left\{\albars^2
-\pi^2\beta_0^2\,\albars^4\right\}+ d_3\,\albars^3 \eeq }
with $\delta_3$ and $\delta_4$ defined\cite{rad82,kat95} in eq.(\ref{deltas}). \par

 The three-- and two--term structures in curly brackets are related to
specific expansion functions \tildal and \agoth defined below (\ref{Rpi})
and entering into the  non-power expansion (\ref{r-new}).\medskip

  To estimate roughly the numerical effect of using this last modified
expression (\ref{4thterm}), we take the case of $e^+e^-$ inclusive
annihilation. For $\sqrt{s}\simeq 3\div5\:\GeV\,$ with
$\albars \simeq 0.28\div 0.22\,$ one has
$$
\left.(\triangle\albars)_4=\frac{\pi\delta_4\,\albars^4}{1+2\pi d_2
\albars}\,\right|_{3\div 5 \GeV}^{f=4} =\frac{1.07\,\albars^4}{1+0.974\,
\albars}\simeq 0.005 \div 0.002 $$
--- an important effect on the level of ca $1 \div 2 \% \,.$

 Moreover, the $(\triangle\albars)_4\,$ correction turns out to be
noticeable even in the lower part of the $f=5\,$ region! Indeed,
at $\sqrt{s} \simeq 10 \div 40 \: \GeV\,$ with $\albars \simeq 0.20 \div
0.15\,$ we have
$$
\left.(\triangle\albars)_4\right|_{10\div 40\:\GeV}^{f=5} \simeq 0.71 \,
\albars^4\simeq (1.1\div 0.3) \cdot 10^{-3} \quad (\lesssim 0.5\%)\,.$$

\section{\sf Non-power expansion in the Minkowskian region \label{s3}}

 The so--called $\pi^2$ terms in the $s$--channel perturbative expansions
for the invariant coupling and observables have a simple origin. \par
 As it is well known, the usual invariant coupling originally defined
\cite{rg56} in terms of real constants $z_i$,
counter-terms of finite Dyson renormalization transformation, can be
expressed via a product of dressed symmetric vertex and propagator amplitudes
taken at space-like values of their arguments.
 $$
\bar\alpha(Q^2,\alpha)=\alpha\Gamma^2(Q^2,\alpha)\prod_i d_i(Q^2,\alpha)\,.$$
 Hence, by construction, it is a real function defined in the Euclidean region.
 \par
  Transition to the time-like region, with logs branching
$\ln Q^2 \to \ln s-i\pi\,$ transforms all relevant amplitudes into complex
functions $\Gamma(s,\alpha), d_i(s,\alpha)\,.$  Here, the problem of
appropriate defining of effective coupling in the time-like domain arises.
\par \smallskip

  For this goal, we shall follow the idea devised in the early 80s
by Radyushkin \cite{rad82} and Krasnikov--Pivovarov \cite{kras82}. There, an
integral transformation $\,{\bf R}$
reverse to the dipole representation for the Adler function has been used.
\par
 We propose {\sf to treat this representation as an integral operation}
\begin{equation}\label{d-trans}
R(s) \to D(z) = Q^2\int^{\infty}_0 \frac{d s}{(s+z)^2}\,R(s)\,
\equiv {\cal\bf D} \left\{ R(s)\right\} \eeq
transforming a function $\,R(s)\,$ of a real positive (time-like) argument
into a function $\,D(z)\,$ given in the cut complex plane with analytic
properties equivalent to those following from the K\"allen--Lehmann
integral representation. In particular, the function $\,D(Q^2)\,$ is real
on the positive (space-like) real axis at $z=Q^2+i0\,; Q^2\geq 0\,.$\par

The reverse operation is expressible in the form of a contour integral
$$
R(s)=\frac{i}{2\pi}\,\int^{s+i\varepsilon}_{s-i\varepsilon}\frac{d
z}{z}\, D_{\rm pt}(-z)\equiv{\bf R}\left[D_{\rm }(Q^2)\right]\,.$$

 With the help of the latter, one can define\cite{js95,ms97} an
effective invariant time-like coupling $\tildal(s)=
{\bf R}\left[\albars(Q^2)\right]\,.$ Omitting some technical details, we
give a few resulting\cite{rad82,kras82,ms97} expressions.\smallskip

E.g., starting with one--loop
$ \albars^{(1)}=\left[\beta_0\ln(Q^2/\Lambda^2)\right]^{-1}\,$
one has ${\bf R}\left[\albars^{(1)}\right]\,$ ---
\beq \label{tildal1}
\tildal^{(1)}(s)=\frac{1}
{\beta_0}\left[\frac{1}{2}-\frac{1}{\pi}\arctan\frac{L}{\pi}\right]_{L>0}=
\frac{1}{\beta_0\pi}\arctan\frac{\pi}{L}\,;\quad L=\ln\frac{s}{\Lambda^2}
\,.\eeq
   At the same time, to $\left(\albars^{(1)}(Q^2)\right)^2\,$ and
$\left(\albars^{(1)}(Q^2)\right)^3\,$  there correspond
$$
\agoth_2^{(1)}(s)\equiv {\bf R}\left[\left(\albars^{(1)}\right)^2\right]
=\frac{1}{\beta_0^2\left[L^2+\pi^2\right]}\,\quad \mbox{and} \quad
\agoth_3^{(1)}(s)=\frac{L}{\beta_0^3\left[L^2+\pi^2\right]^2} \,. $$
 In the two--loop case, for a ``popular" expression
$$\beta_0\bar{\alpha}_{s,pop}^{(2)}(Q^2)=\frac{1}{l}-b_1(f)
\frac{\ln l}{l^2}\:; \quad l=\ln\frac{Q^2}{\Lambda^2}\,$$
 one obtains\cite{rad82} the two-loop ``pop" effective $s$--channel coupling
\beq\label{tildal2pop}
\tildal^{(2)}_{pop}(s)=\left(1+\frac{b_1 L}{L^2+\pi^2}
\right)\tildal^{(1)}(s) -\frac{b_1}{\beta_0}\frac{\ln\left[
\sqrt{L^2+\pi^2}\right]+1}{L^2+\pi^2}\,. \eeq
 Both the expressions (\ref{tildal1}) and (\ref{tildal2pop}) are
monotonically decreasing with a finite IR $\,\tildal(0)=1/\beta_0(f=3)\simeq
1.4\,$ value. Meanwhile, higher functions go to the zero $\agothk(0)=0\,$
at the IR limit. \par

In the case $L\gg \pi$, it is possible to expand \tildal and $\agoth_k$
in powers of $\pi^2/L^2$. Then functions \tildal and $\agoth_2$ can be presented
as expansions in powers of common $\albars \simeq 1/L$. They correspond to curly
brackets in (\ref{4Delta}).\par\bigskip

 In \cite{rad82,kras82}, as a starting point for observables in the
Euclidean, i.e., space-like domain $\,Q^2 >0$, the perturbation series
$$
D_{\rm pt}(Q^2)=1+ \sum_{k\geq 1}^{} d_k\,\albars^k(Q^2) $$
has been assumed. It contains powers of usual, RG summed, invariant
coupling $\albars(Q^2)\,$ that obeys unphysical singularities in the
infrared (IR) region around $Q^2 \simeq \Lambda^2_3\,$. \par

 By using the ${\bf R}$ transformation, we obtain in the Minkowskian
region the ``transformed" expansion over a non-power set of functions
\begin{equation}\label{Rpi}
R_{\pi}(s)\equiv {\bf R}\left[D_{\rm pt}(Q^2)\right]=1+\sum_{k\geq 1}d_k
{\mathfrak{A}}_k(s)\,;\quad ~\mathfrak{A}_k(s)={\bf R}\left[\albars^k(Q^2)
\right]\,\eeq
free of the mentioned singularities. Properties of these functions have
been analyzed in detail in our previous paper\cite{tow00} --- see also
Ref. \cite{dec00}. For a more detailed numerical information on the
functions \tildal, $\agoth_2$ and $\agoth_3$ see Ref.\cite{mag00}. \par
Here, we give condensed information
that will be enough for a few illustrations.  \smallskip

\begin{minipage}[t][]{125mm}
\begin{center}{\sf\large Table 2 ~} \\ \mbox{\bf Three-loop APT results
for $\Lambda_{\rm \msbar}^{(5)}=290\:\GeV \,; ~\asmzs=0.125$} \medskip

\begin{tabular}{|c||c|c|c|c|c|c|c|c|c|}  \hline
$\sqrt{s}/\GeV$& 5 & 10 &15    & 20  & 30  & 50 & 60 & 90  & 150 \\ \hline\hline
$\albars(s)$&.235&.195  &.177  &.165 &.153 &.137 &.133 &.125 & .115 \\ \hline
$\tildal(s)$&.221&.186&.170 &.160&.148&.136&.132&.123&.114 \\
$10\agoth_2$&.456 &.330&.275 &.246&.214&.180&.169&.149& .129 \\
$100\agoth_3$&.871&.555 &.436  &.357 &.299 &.232&.213&.177 &.143  \\ \hline
\end{tabular} \end{center}
\end{minipage}
\bigskip

 Both in the Figure 1 and in Table 2, we give 3-loop solutions
for \albars  as well as for the modified, so--called {\it global}
(for detail, see paper \cite{tow00}) functions
$\tildal=\agoth_1\,, \ \agoth_2$ and $\agoth_3$ calculated within
the \msbar scheme for the cases $\Lambda_{(5)}=
215\,\GeV\,,~\asmz=0.118 \,$ and
 $\, \Lambda_{(5)}= 290\,\GeV\,$, $\,\asmz=0.125\,.$ \par \smallskip

   \begin{figure}[h]\label{fig-1}
\unitlength=1mm
\begin{picture}(0,100)
  \put(-20,-5){%
 \epsfig{file=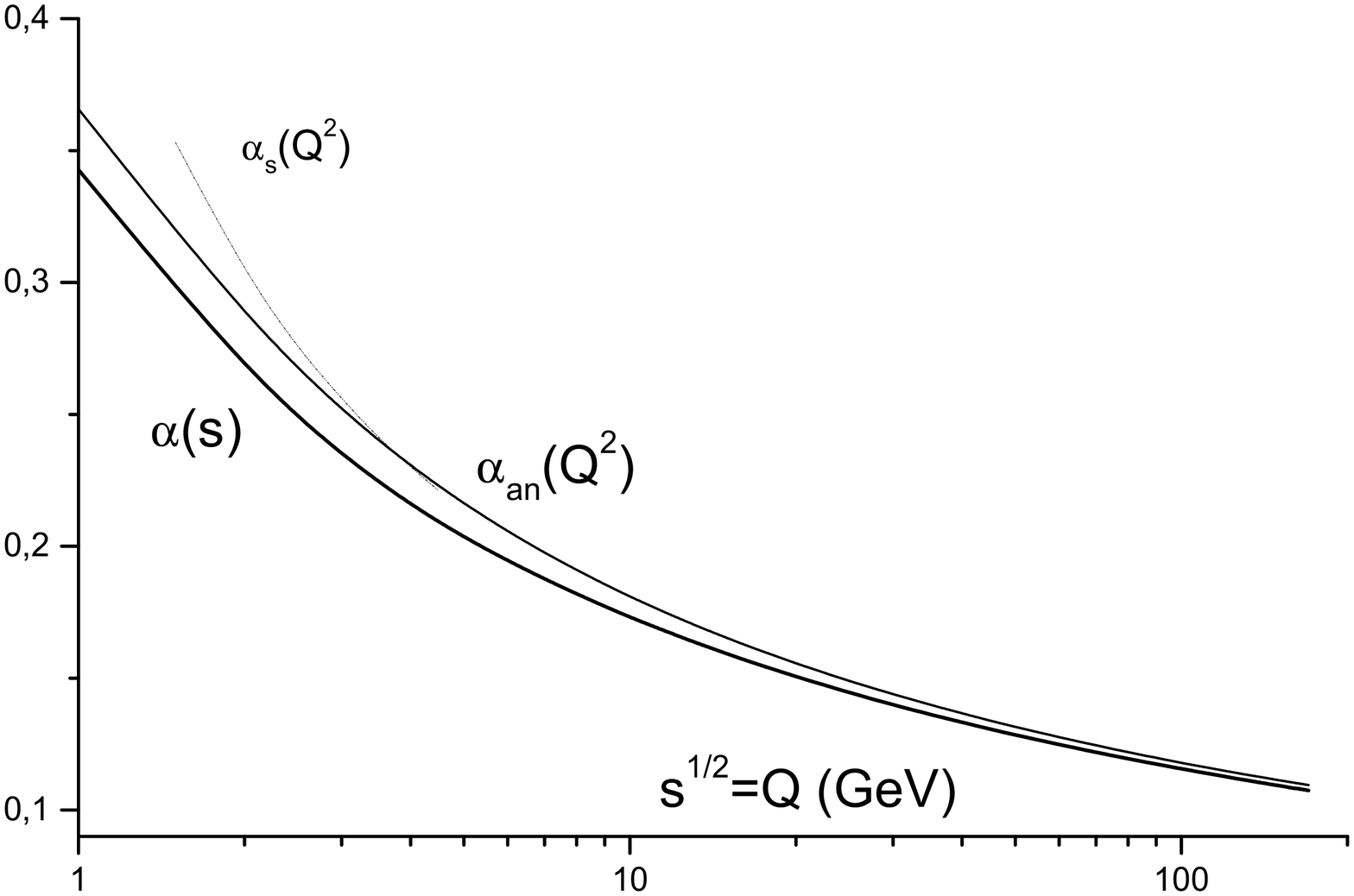,width=160mm,height=100mm}%
}
\end{picture}
 \caption{\sl  Effective global Minkowskian, \tildal, and Euclidean,
\alphan expansion functions, as \hspace{1mm} compared with the
standard one \albars (at $\Lambda_{(5)}=350 \,\MeV$ and
$~\asmz=0.118$).}
 \end{figure}

We have chosen these two cases as limiting ones as far as in many
practical cases real figures lie between these limits. \par \smallskip

  In the first figure we give three curves \albars, \tildal and \alphan
related to the same physical case for $\Lambda_3=350\,\MeV$ and $\asmz=0.118$.
 The curves \tildal and \alphan on the figure go a bit
slanting than usual, the \albars, dotted curve. This is quite natural, as
they both are regular in the vicinity of the $\Lambda$ singularity.  \par

Meanwhile, only two first, \tildal and \alphan have direct
physical meaning (compare with conclusion of \cite{tow00}). Just
their values have to be determined from any given experiment.
Nevertheless, in the four- and five--flavour regions one can still
refer to \albars and \asmz as to traditional theoretical objects. \par
\smallskip

 Now, instead of (\ref{rnew}), with due account to (\ref{Rpi}), we have
\beq\label{r-new}
r(s)=\frac{\tildal(s)}{\pi}+d_2\,{\agoth}_2(s)+d_3\:\agoth_3(s)\:\eeq
 with beautifully decreasing coefficients $\,d_k\,.$
Just this nonpower expansion, strictly speaking, should be used
instead of its approximations, eqs.(\ref{3Delta}) and
(\ref{4Delta}), for data analysis in  the time-like region. \par

 At the same time, in the Euclidean, we have also non-power expansion
\beq\label{d-new}
d(Q^2)=\frac{\alphan(Q^2)}{\pi}+d_2\,\acal_2(Q^2)+d_3\:\acal_3(Q^2)\:\eeq
that can be related to (\ref{r-new}) by transformation (\ref{d-trans}) in
the framework of Invariant Analytic Approach (refs.\cite{prl97,ss99tmp}).\par

 These non-power expansions, free of unphysical singularities, jointly form
a correlated system. The latter has been studied in detail in Refs.\cite{tow00} and
\cite{apt00}. We call it Analytic Perturbation Theory (APT).

\section{\sf Numerical illustrations}

 To illustrate, let us start with a few cases in the $f=5\,$ region. \par \medskip

To begin with, consider \underline{\sf the $\Upsilon$ decay.} According to
the Particle Data Group (PDG) overview (see their Fig.9.1 on page 88 of
Ref.\cite{pdg00}), this is (with $\as(M_{\Upsilon})\simeq 0.170$ and
$\asmz=0.114$) one of the most ``annoying" points of their summary of
\asmz values. It is also singled out theoretically. The expression for
the ratio of decay widths starts with the cubic term
\beglab{up-wid}
R(\Upsilon)=R_0\,\as^3(M_{\Upsilon})(1+e_1\,\as)\quad \mbox{with}\quad
e_1\simeq 1\,.\eeq \par

Due to this, the $\pi^2$ correction\footnote{First proposal of taking
into account this effect in the $\Upsilon$ decay was discussed\cite{kras82}
more than a quarter of century ago. Nevertheless, in current
practice it is neglected.} is rather big here
\beglab{ag3}
  \agoth_3 \simeq \as^3\left(1- 2(\pi\beta_0)^2 \as^2 \right)\,.\eeq
Accordingly,
$$\Delta\as(M_{\Upsilon})=\frac{2}{3}\,(\pi\beta_0)^2\,\as^3(M_{\Upsilon})
\simeq 0.0123 \,,$$ 
that corresponds to
\beglab{del-ups-asz}
\Delta\as(M_{Z}) = 0.006 \quad \mbox{with} \quad \as(M_{Z}) = 0.120\,. \eeq

  Now, let us turn to a few cases analyzed by the three-term expansion formula
(\ref{rnew}). For the first example, take \underline{\sf $e^+e^-$ hadron
annihilation} at $\sqrt{s}=42\,\GeV$ and $11\,\GeV\,.$

A common form (see, e.g., Eq.(15) in Ref.\cite{beth00}) of
theoretical presenting of the QCD correction in our normalization
looks like
\beq\label{R-Z}
r_{e^+e^-}(s)=0.318\albars(s) + 0.143\,\albars^2 -0.413\,\albars^3\,.
\eeq
 Starting with $r_{e^+e^-}(42)\simeq 0.0476$, one has
$\albars(42)=0.144\,.$ Along with our new philosophy, one should use
instead
\beq\label{r-annih}
r_{e^+e^-}(s)=0.318\,\tildal(s)+0.143\,\agoth_2(s)-0.023\,\agoth_3(s)\eeq
that yields $\tildal(42)=0.142\,$ with $\as(42)=0.145$ and $\asmz=0.127\,$
to be compared with $\asmz=0.126\,$ under a usual analysis. \par \smallskip

 Quite analogously, for $r_{e^+e^-}(11)\simeq 0.0661\,;\:\albars(11)=
0.200\,,$ we obtain $\tildal(10)=0.190\,$ that corresponds to
$\asmz=0.129$ instead of 0.130. \medskip

  For the next example, we take the \underline{\sf $\/Z_0$ inclusive decay}.
Experimental ratio $R_Z=\Gamma(Z_0\to hadrons)/\Gamma(Z_0\to leptons)=20.783
\pm.029\,$ is usually presented as follows: $R_Z=R_0\left(1+r_Z(M_Z^2)\right)\,$
with $R_0=19.93\,.$ A common form (see, e.g., Eq.(15) in Ref.\cite{beth00})
of presenting of the QCD correction in our normalization looks like
$$
r_Z(M_Z^2)=0.3326\albars + 0.0952\,\albars^2 -0.483\,\albars^3\,.$$

  To $\left[r_Z\right]_{obs}= 0.04184\:$ there corresponds
$\:\asmz=0.1241\,$ with $\,\Lambda_{\msbar}^{(5)}=292 \,\MeV\,.$
In the APT case,from
\beq\label{r-pi}
r_Z(M_Z^2)=0.3326\,\tildal(M_Z^2)+0.0952\,\agoth_2(M_Z^2)-0.094\,
   \agoth_3(M_Z^2)\eeq
we obtain $\tildal(M_Z^2)= 0.122$ and $\:\asmz=0.124\,$ that relates to
$\Lambda^{(5)}=290 \,\MeV\,.$
Note that here the three-term approximation of (\ref{4Delta}) gives the same
relation between the \asmz and $\:\tildal(M^2_Z)\,$ values. \smallskip

 Nevertheless, in  accordance with our preliminary estimate for
the $\,(\triangle\albars)_4\,$ role, even the so-called NNLO theory needs
some  $\pi^2$ correction in the $W=\sqrt{s}\lesssim 50\:\GeV$ region.
\smallskip

 Now,  turn to the experiments in the HE Minkowskian (mainly with a shape
analysis) that usually are confronted with two-term  expression
(\ref{two-term}). As it has been shown below, the main theoretical error
in the $\,f=5\,$ region can be expressed in the form
\beglab{das5}
\left.(\triangle\albars(s)\right|_{20\div 100 \GeV}^{f=5}\simeq
1.225\,\albars^3(s)\simeq0.002\div0.003\,.\eeq
An adequate expression for the shift of an equivalent $\asmz$ value is
\beglab{dasmz}
[\triangle\asmz]_3=1.225\albars(s)\asmz^2\,.\eeq  %
\smallskip

We give results of our approximate APT calculations, mainly by
Eqs.(\ref{das5}) and (\ref{dasmz}), in the form of Table 3 and Figure 2. At
the last column of the Table 3 in brackets we indicate difference between
the APT and usual analysis. By bold figures the results of the three--loop
analysis are singled out.

\begin{minipage}[t]{130mm} 
\begin{center}
{\sf\large Table 3 \label{tab3}} \smallskip

{\bf The APT revised\footnote{``j \ \& \ sh" = jets and shapes;
Figures in brackets in the last column give the \\ dif\-ference
$\Delta\asmz$ between common and APT values.}
part ($f=5$) of Bethke's\cite{beth00} Table 6 } \medskip

\begin{tabular}{|c|c|c||c|c||c|c|}  \hline \smallskip

&$\sqrt{s}$&loops&\albars(s)&\asmzs&\albars(s)&\asmzs\\ Process&\GeV& No&
ref.[2] &ref.[2]& APT &APT  \\ \hline \hline
$\Upsilon$-decay \footnote{Taken from Ref.\cite{pdg00}.}&9.5&2 &.170&.114 &.182&.120 (+6)         \\
$e^+e^-[\sigma_{had}]$&10.5&{\bf 3}&.200&.130 &.198&.129({\bf -1}) \\ 
$e^+e^-[j\, \&\, sh]$   &22.0&2 &.161&.124 &.166&.127(+3)  \\
$e^+e^-[j\, \&\, sh]$   &35.0&2 &.145&.123 & .149&.126(+3)  \\ 
$e^+e^-[\sigma_{had}]$&42.4&{\bf 3} &.144&.126 &.145&.127(+{\bf1}) \\ 
$e^+e^-[j\, \& sh]$     &44.0&2 &.139&.123 &.142&.126(+3) \\ 
$e^+e^-[j\, \& sh]$     &58  &2 &.132&.123 &.135&.125(+2)  \\ 
{\bf $Z_0\to$ had.}     &91.2&{\bf 3} &.124&.124 &.124&.124 ({\bf 0})  \\
$e^+e^-[j\,\&\,sh]$     &91.2&2 &.121&.121&.123&.123(+2) \\
$e^+e^-[j\,\&\,sh]$     &133 &2 &.113&.120& .115&.122(+2)\\
$e^+e^-[j\,\&\,sh]$     &161 &2 &.109&.118& .111&.120(+2) \\
$e^+e^-[j\,\&\,sh]$     &172 &2 &.104&.114& .105&.116(+2)\\
$e^+e^-[j\,\&\,sh]$     &183 &2 &.109&.121& .111&.123(+2) \\
$e^+e^-[j\,\&\,sh]$     &189 &2 &.110&.123& .112&.125(+2) \\ \hline
\end{tabular}\end{center}
Averaged $<\asmzs>_{f=5}$ values \hspace{15mm}\/$0.121;$
\hspace{18mm} $0.124\,;$
\end{minipage}
\vspace{10mm}\bigskip

 Let us note that our average over events from Table 6 of Bethke's review
\cite{beth00} nicely correlates with recent data of the same author (see
Summary of Ref.\cite{beth-fest}). The best $\chi^2$ fit yields $\asmz_{[2]}
=0.1214$ and $\asmz_{APT}=0.1235\,.$ This gives minimum $\chi^2_{[2]}=0.197$
and $\chi^2_{APT}=0.144\,$ with impressive ratio ($\simeq 0.73$)
illustrating the effectiveness of the APT procedure.\par

 On the Fig.2 by open circles and bullets $(\circ, \bullet)$  we give two--
and three--loops data mainly from Fig.10 of paper \cite{beth00}. The only
exclusion is the $\Upsilon$ decay taken from the Table 6 of the same paper.
By crosses we marked the new ``APT values" calculated approximately mainly
with help of Eq.(\ref{das5}).

For clearness of the \pisq effect, we skipped the error bars. They are the same
as in the mentioned Bethke's figure and we used them for calculating $\chi^2\,.$
   \begin{figure}[h]\label{fig-2}
\unitlength=1mm
\begin{picture}(0,100)
  \put(3,-5){%
   \epsfig{file=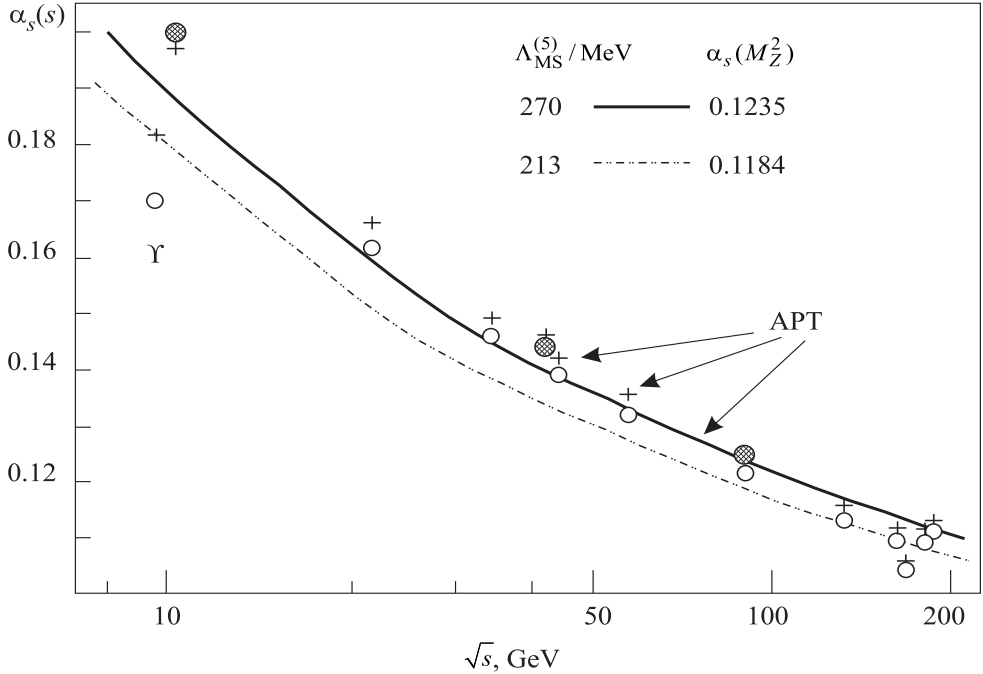,width=130mm,height=90mm}%
}
\end{picture}
\caption{\small The new APT analysis for \albars in the five-flavour time--like
region. Crosses (+) differ from circles $(\circ, \bullet)$ by $\pi^2$ correction
(\ref{das5}). Solid APT curve relates to $\Lambda_{\rm MS}^{(5)}=270 \,\MeV$ and
$~\asmz=0.124\,.$ To compare, we give also the standard
(dot-and-dash curve) \albars (at $\Lambda^{(5)}=213 \,\MeV$ and
$~\asmz=0.118$) taken from Fig.10 of paper \cite{beth00}.}
 \end{figure}

\newpage
\section{\sf Conclusion} 
 We have established a few qualitative effects: \par

{\bf 1.} Effective positive shift $\Delta\albars = +0.002\,$ in the upper
 half ($\geq \, 50 \,\GeV$) of the $f=5\,$ region for all time-like
events that have been analyzed up to now in the NLO mode.\par\smallskip

{\bf 2.} Effective shift $\Delta\albars \simeq +0.003\,$ in the lower half
($10 \div 50 \, \GeV$) of the $f=5\,$ region for all time-like
events that have been analyzed in the NLO modes.\par\smallskip
{\bf 3.} The new value
\beq\label{124}
 \asmz=0.124\, \eeq
by averaging over the $f=5$ region.
\par\medskip

  These results are based on a plausible hypothesis on the ``$\pi^2\,$--
terms" prevalence in expansion coefficients for observable in the
Minkowskian domain. The hypothesis has some preliminary support but
needs to be checked in a more detail. \par
 Nevertheless, our result (\ref{124}) being taken as granted, rises
 two physical questions: \par
 -- The issue of self-consistency of QCD invariant coupling behavior
between the ``medi\-um $(f=3,4)$" and ``high  $(f=5,6)$" regions. \par
 -- The new ``enlarged value" (\ref{124}) can influence various
physical speculations in the several hundred \GeV \,region. \bigskip

{\bf\large Acknowledgements} \smallskip

 The author is indebted to D.Yu.~Bardin, N.V.~Krasnikov, B.A.~Magradze,
S.V. Mi\-k\-hai\-lov, A.V. Radyushkin, I.L. Solovtsov and O.P. Solovtsova
for useful discussions and comments. This work was partially supported by
grants of the Russian Foundation for Basic Research (RFBR projects Nos
99-01-00091 and 00-15-96691), by INTAS grant No 96-0842 and by INTAS-CERN
grant No 2000-377.


\begin{thebibliography}{70}
\bibitem{pdg00} D.E.Groom {\it et al.}, {\it European Phys. J.} {\bf C 15}
                       (2000) 1.
\bibitem{beth00} S. Bethke, ``Determination of the QCD coupling $\alpha_s$",
               {\it J. Phys.} {\bf G 26} R27; Preprint MPI-PhE/2000-07,
                             April 2000; hep-ex/0004021.
\bibitem{bardin} D.Yu. Bardin, G. Passarino, {\sf The Standard model               
                       in the making}, Clarendon Press, Oxford, 1999.
\bibitem{rad82} A. Radyushkin, Dubna JINR preprint E2-82-159 (1982); see also
          {\it JINR Rapid Comm.} No.4[78]-96 (1996) pp 9-15 and hep--ph/9907228.
\bibitem{kras82} N.V. Krasnikov, A.A. Pivovarov,
             {\it Phys. Lett.} {\bf 116 B} (1982) 168--170.
\bibitem{bjork89} J.D. Bjorken, ``Two topics in QCD", Preprint SLAC-PUB-5103
           (Dec 1989); in {\sf Proc. Cargese Summer Institute}, eds. M. Levy
           et al., Nato Adv. Inst., Serie B, vol. 223, Plenum, N.Y., 1990.
\bibitem{kat95} A.L. Kataev and V.V. Starshenko, {\it Mod. Phys. Lett. A}
                {\bf 19} (1995) 235-250.
\bibitem{delphi} Delphi Collaboration, `` Consistent measurements of
                  \as from ...", CERN-EP/99-133 (Sept 1999).
\bibitem{rg56} N.N. Bogoliubov and D.V. Shirkov, {\sl Doklady AN SSSR},
        {\bf 103} (1955) 203-206 (in Russian); also {\it Nuovo Cim.}
   {\bf 3} (1956) 845-637; {\it Sow. Phys. JETP\/} {\bf 3} (1956)  57--64
         and Chapter ``Renormalization group" in \cite{kniga}.
\bibitem{kniga} N.N. Bogoliubov and D.V. Shirkov, {\sf Introduction to the
         theory of Quantized Fields}, Wiley \& Intersc. N.Y., 1959 and 1980.
\bibitem{js95} H.F.~Jones and I.L.~Solovtsov, {\it Phys. Let.} {\bf B 349}
               (1995)  519-525.
\bibitem{ms97} K.A.~Milton and I.L.~Solovtsov, {\it Phys. Rev.} {\bf D 55}
               (1997)  5295-5298.
\bibitem{tow00} D.V. Shirkov, ``Toward the correlated analysis of observables
             in perturbative QCD", JINR preprint E2-2000-46; hep-ph/0003242.
\bibitem{dec00} D.V. Shirkov, ``Analytic Perturbation Theory for QCD
                 observables", JINR preprint E2-2000-298; hep-ph/0012283
\bibitem{mag00} B. Magradze, ``The QCD coupling up to third order in standard
                 and analytic perturbation theories", Preprint JINR,
                   E2-2000-222; hep-ph/0010070;
\bibitem{prl97} D.V. Shirkov and I.L. Solovtsov, {\it Phys.Rev.Lett.}
                     {\bf 79} (1997) 1209-12; hep-ph/9704333.
\bibitem{ss99tmp} I.L. Solovtsov and D.V. Shirkov, {\it Theor. Math. Phys.}
               {\bf 120} (1999) 1210--1243; hep-ph/9909305.
\bibitem{apt00} D.V. Shirkov, ``Analytic Perturbation Theory for QCD observables",
                         JINR preprint E2-2000-296; hep-ph/0012283.
\bibitem{beth-fest} S. Bethke, ``Standard Model Physics at LEP", hep-ex/0001023
                   and ``QCD at LEP", (Talk at LEP Fest on Oct 11, 2000).
\end{thebibliography}
\end{document}